# Formation of Glycerol through Hydrogenation of CO ice under Prestellar Core Conditions

G. Fedoseev[1,2], K.-J. Chuang[1,3], S. Ioppolo[4], D. Qasim[1], E.F. van Dishoeck[3], and H. Linnartz[1]

[1]Sackler Laboratory for Astrophysics, Leiden Observatory, Leiden University, PO Box 9513, NL-2300 RA Leiden, the Netherlands

[2]INAF – Osservatorio Astrofisico di Catania, via Santa Sofia 78, 95123 Catania, Italy; gfedo@oact.inaf.it

[3]Leiden Observatory, Leiden University, PO Box 9513, NL-2300 RA Leiden, the Netherlands

[4]School of Physical Sciences, The Open University, Walton Hall, Milton Keynes MK7 6AA, UK

## ABSTRACT

Observational studies reveal that complex organic molecules (COMs) can be found in various objects associated with different star formation stages. The identification of COMs in prestellar cores, *i.e.*, cold environments in which thermally induced chemistry can be excluded and radiolysis is limited by cosmic rays and cosmic ray induced UV-photons, is particularly important as this stage sets up the initial chemical composition from which ultimately stars and planets evolve. Recent laboratory results demonstrate that molecules as complex as glycolaldehyde and ethylene glycol are efficiently formed on icy dust grains *via* non-energetic atom addition reactions between accreting H atoms and CO molecules, a process that dominates surface chemistry during the 'CO-freeze out stage' in dense cores. In the present study we demonstrate that a similar mechanism results in the formation of the biologically relevant molecule glycerol – $HOCH_2CH(OH)CH_2OH$ – a three-carbon bearing sugar alcohol necessary for the formation of membranes of modern living cells and organelles. Our experimental results are fully consistent with a suggested reaction scheme in which glycerol is formed along a chain of radical-radical and radical-molecule interactions between various reactive intermediates produced upon hydrogenation of CO ice or its hydrogenation products. The tentative identification of the chemically related simple sugar glyceraldehyde – $HOCH_2CH(OH)CHO$ – is discussed as well. These new laboratory findings indicate that the proposed reaction mechanism holds much potential to form even more complex sugar alcohols and simple sugars.

## 1. INTRODUCTION

With the increasing sensitivity of astronomical observing facilities, more and more complex organic molecules, so called COMs, have been identified in space (Halfen et al. 2015; Jørgensen et al. 2016; Cernicharo et al. 2016; McGuire et al. 2016). Gas phase reactions are not efficient enough to explain many of the observed COM abundances and it is generally accepted that such species form on the surfaces of icy dust grains by merging smaller species to larger and larger constituents. This idea, supported by numerous laboratory results and astrochemical simulations, indicates that surface reactions, triggered by accreting atoms, impacting cosmic rays and VUV-photons, as well as by thermal processing provide efficient pathways towards molecular complexity (Charnley et al. 2008; Garrod et al. 2008; Herbst & van Dishoeck 2009; Vasyunin & Herbst 2013; Walsh et al. 2014a,b; Linnartz et al. 2015; Öberg et al. 2016).

Key in the formation of new COMs is the mechanism that extends, for interstellar conditions, the size of a carbon backbone of the precursor molecule with new C-C bonds. Subsequently, these larger species act as parents in new reactions, resulting in the formation of a whole set of other COMs with similar size properties. This holds until another C-C bond is added to one of these reaction products, yielding a parent species that acts as precursor for another set of even larger COMs, *etc.*. This repeatable process can be best illustrated by the reaction scheme introduced by Charnley et al. (2001, 2008). During the 'catastrophic CO freeze-out' in dense cores, accreting CO molecules produce a layer of CO ice on the icy grain surfaces. Simultaneously, CO molecules undergo partial hydrogenation by accreting H-atoms, producing $CH_3OH$ in a series of H-atom addition and abstraction reactions that start with the formation of HĊO radicals (Hiraoka et al. 1994; Zhitnikov & Dmitriev 2002; Watanabe & Kouchi 2002; Fuchs et al. 2009; Hidaka et al. 2009). Such 'non-energetic' atom addition reactions, *i.e.* reactions between species that are in thermalized equilibrium with the grain surface and, therefore, not able to cross large reaction barriers, play a main role in dark cloud chemistry. The subsequent addition of C-atoms to HĊO results in the formation of ·CCHO and ·CCCHO radicals. The latter two intermediates are, in turn, proposed starting points in the formation of entire sets of new COMs through a sequence of H- and O-atom addition reactions. For example, acetaldehyde, ethanol, glyoxal, glycolaldehyde, and ethylene glycol are all produced through a series of atom or molecule additions to ·CCHO radicals. In a similar way, their 'three-carbon bearing' analogues are produced from ·CCCHO intermediates.

The chemical concept of this C-atom addition scenario is straight forward, but requires a long sequence of 'site-selective' H-, C- and O-atom additions, making efficient formation of species like glyceraldehyde and ethylene glycol (in which each of the carbon atoms has one oxygen atom attached) unlikely. Moreover, in dense clouds a most of the atomic carbon is locked in CO molecules, and fewer carbon atoms are available for accretion on the icy grain surface. Therefore, the formation of these two species has been addressed in another scenario that was experimentally verified by Fedoseev et al. (2015) and later extended by Butscher et al. (2015) and Chuang et al. (2016). In this reaction scheme, a carbon backbone increase is realized through recombination of various C(O)-bearing intermediate radicals that are produced along the CO to CH$_3$OH hydrogenation route (see Figure 8 of Chuang et al. 2016). It was recently proven that direct recombinations of HĊO and ·CH$_2$OH free radicals form glycolaldehyde, ethylene glycol and, possibly, glyoxal. In this scheme, ethylene glycol may also be formed *via* two sequential H-atom additions to glycolaldehyde.

In an alternative 'energetic processing' scenario, reactive intermediates are formed upon irradiation of CH$_3$OH-rich ices by various energetic particles including protons, electrons, X-rays, and VUV-photons, *etc.*. Laboratory studies show that in this case COMs are efficiently formed through recombination of carbon bearing CH$_3$OH dissociation products (Allamandola et al. 1988; Hudson & Moore 2000; Öberg et al. 2009; Modica & Palumbo 2010; Jheeta et al. 2013; Henderson & Gudiparti 2015; Maity, Kaiser & Jones 2015; Abou Mrad et al. 2016; Paardekooper et al. 2016; Chuang et al. 2017). Whereas the 'non-energetic' scenarios are thought to dominate in dark clouds, the latter ones will be more prominent during later stages of the star and planet formation process.

Efficient mechanisms increasing the number of C-C bonds are particularly interesting from an astrobiological point of view. Many of the basic chemical compounds required for the emergence of life consist of chains of carbon atoms with repeatable functional groups, such as (un)saturated fatty acids, simple sugars and sugar alcohols. The two latter classes deserve special attention, as their smallest members, glycolaldehyde (HOCH$_2$CHO) and ethylene glycol (HOCH$_2$CH$_2$OH), have been successfully observed in the gas phase toward our galactic center (Hollis, Lovas & Jewell 2000; Hollis et al. 2002), solar-mass protostars (Jørgensen et al. 2012 and 2016; Taquet et al. 2015; Coutens et al. 2015; Rivilla et al. 2016) and in cometary ices (Biver et al. 2014; Le Roy et al. 2015; Goesmann et al. 2015). It is possible that the mechanism responsible for the formation of glycolaldehyde and ethylene glycol also holds the potential to form the next larger members in these series, glyceraldehyde and glycerol. These molecules are needed in the synthesis of other sugars, sugar alcohols, amino acids and even a nucleic acid, as proposed in one of the possible early Earth geo-chemical scenarios (Ritson & Sutherland 2012; Patel et al. 2015; Sutherland 2017). Thus, studying possible formation routes of this type of COMs for conditions relevant to star forming regions, at stages prior to the formation of planets is particularly intriguing, assuming that at least a fraction of the original icy-dust material survives upon transfer to early planet surfaces.

## 2 PROPOSED REACTION SCHEME

In this work we experimentally verify an extension of the scenario in which COMs are formed through recombination of reactive carbon bearing radicals produced along the CO to CH$_3$OH hydrogenation route. It illustrates that the mechanism resulting in the formation of glycolaldehyde and ethylene glycol is also capable to form larger molecules with a (-C(O)-)$_n$ backbone, *i.e.* glyceraldehyde and glycerol (see **Figure 1** for more details). Glycerol (HOCH$_2$CH(OH)CH$_2$OH) is one of the homologues in the sugar alcohol row, a subclass of linear polyols that can be described by the formula HOCH$_2$(CH(OH))$_n$CH$_2$OH. This class of chemical compounds comprises a chain of carbon atoms where each of the carbons is linked to the single hydroxyl (OH) group. With n=0 this formula is reduced to ethylene glycol, virtually the first representative in this row. The significant structural similarity between glycerol and ethylene glycol (the final product of glycolaldehyde hydrogenation) complicates the successive assignments of glycerol by means of IR spectroscopy, as the vibrational modes of glycerol are nearly identical to those of ethylene glycol and, to some extent, of CH$_3$OH. Broadening of absorption features due to hydrogen bonds, further complicates spectroscopic assignments. In a similar way glyceraldehyde links to the series HOCH$_2$(CH(OH))$_n$CHO. In order to circumvent these problems, we use a combination of temperature programmed desorption and quadrupole mass spectrometry (QMS). The methodology is described in the next section.

The experiments presented below are aimed to reproduce the 'CO-freeze out' stage in dense cores (Tielens et al. 1991; Pontoppidan 2006). At this stage external UV-fields are to a large extent shielded by dust grains. This results in a significantly higher rate of H-atom accretion events on an ice surface in comparison to the number of impacting VUV-photons or cosmic rays, *e.g.* H-atom flux ~$10^4$ atoms cm$^{-2}$ s$^{-1}$ *vs.* UV-flux of ~$1$-$10\times10^3$ photons cm$^{-2}$ s$^{-1}$ (Prasad & Tarafdar 1983, Chuang et al. 2017). Therefore, the CO hydrogenation comprises the core of the reaction scheme presented in **Figure 1**. CH$_3$OH ice has been observed in dense clouds (Pontoppidan 2003; Bottinelli et al. 2010; Boogert et al. 2015), and its abundance is in line with the predictions of theoretical and modeling work (Tielens & Hagen 1982; Shalabiea & Greenberg 1994; Geppert et al. 2005; Cuppen et al. 2009, 2011; Vasyunin & Herbst 2012). Recent experimental and theoretical studies demonstrated that two-carbon bearing COMs, *i.e.*, glycolaldehyde and ethylene glycol, are aslo formed at this stage (Woods et al. 2012; Fedoseev et al. 2015; Butscher et al. 2015; Chuang et al. 2016). The key point in the formation of these species appears to be the recombination of free intermediate radicals, *i.e.*, HĊO or ·CH$_2$OH, produced in the sequence of H-atom addition and abstraction reactions along the aforementioned CH$_3$OH formation route (see Figure 8 of Chuang et al. 2016 for more details). These experimentally verified reaction pathways comprise the left part of the reaction scheme presented in **Figure 1**, and, can be further extrapolated to form even larger species.

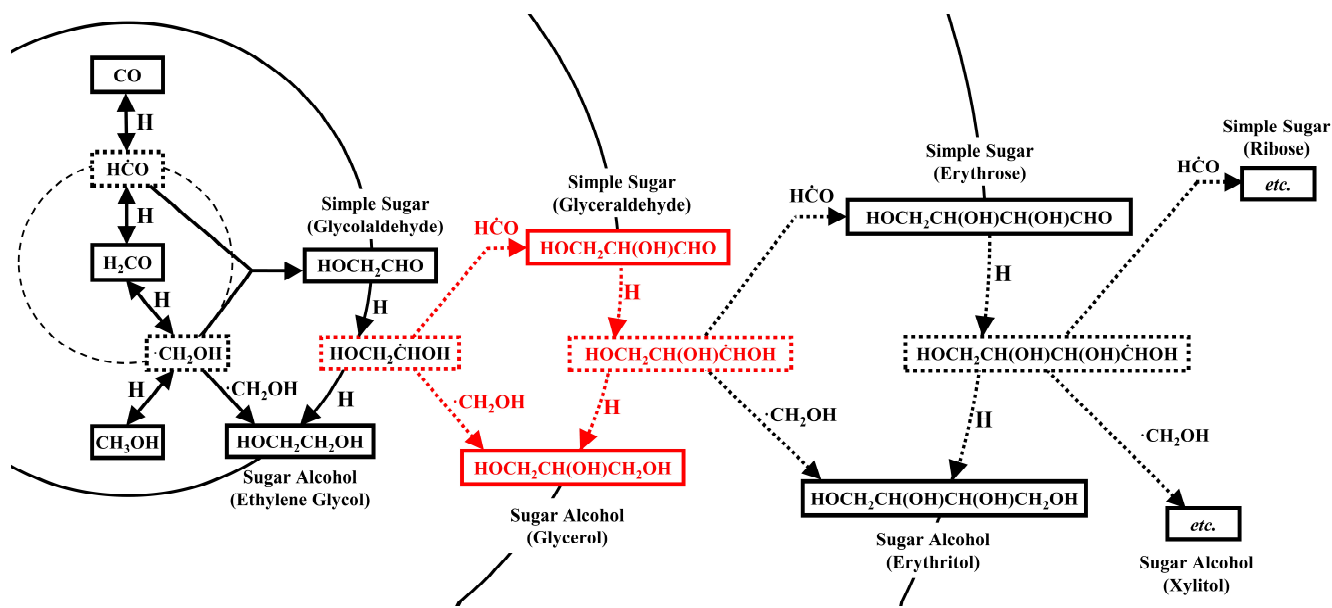

**Figure 1.** Proposed reaction scheme demonstrating the repetitive formation of various simple sugars and sugar alcohols starting from hydrogenation of CO molecules on the surface of interstellar grains. Solid line boxes represent stable molecules, while the intermediate radicals are shown in the dashed line boxes. In the case of more than one possible structural isomer, only the relevant one is depicted for simplicity, *e.g.* both ·CH$_2$OH and CH$_3$O· are formed, but only ·CH$_2$OH is relevant for this scheme, *etc*. Solid arrows present the reaction pathways discussed in previous experimental studies. The dashed arrows show reaction routes made by extrapolation of these previously obtained results. The reaction routes investigated in the present study, resulting in the formation of glycerol and glyceraldehyde, are highlighted in red. Stereochemistry is omitted for simplicity.

Both glycolaldehyde and ethylene glycol can participate in similar H-atom induced reactions, producing two-carbon bearing reactive intermediates, *e.g.* HOCH$_2$ĊHOH. This radical, in turn, can participate in a barrierless recombination with HĊO or ·CH$_2$OH resulting in the formation of three-carbon bearing species, *i.e.*, glyceraldehyde and glycerol, the next simple sugar and sugar alcohol representatives in the HOCH$_2$(CH(OH))$_n$CHO and HOCH$_2$(CH(OH))$_n$CH$_2$OH rows, respectively. These formation routes are illustrated in the central part of **Figure 1** and are experimentally verified in the present study. In an identical way, the same reaction sequence can be further extrapolated to form four-, five- and more-carbon bearing representatives. Here, the consecutive increase of the (-C(O)-)$_n$ backbone by one segment is achieved each time when a HĊO or ·CH$_2$OH radical reacts with a three-, four- or more-carbon bearing precursor, as illustrated in the right part of the reaction scheme. In the figure solid lines represent experimentally verified reaction routes, while dashed lines show their further extrapolation as proposed in this study.

It is important to note that this reaction scheme explains the formation of simple sugars and sugar alcohols already at the cold dark cloud stage, *i.e.*, earlier than in most other suggested chemical scenarios which require thermal heating or irradiation by photons and cosmic rays originating from the newly formed protostar (Garrod & Herbst 2006; Maity, Kaiser & Jones 2015; Meinert et al. 2016; Paardekooper et al. 2016a; Taquet et al. 2016).

## 3. EXPERIMENTAL

The experiments are performed using SURFRESIDE$^2$, an ultra-high vacuum (UHV) setup that has been described in much detail by Ioppolo et al. (2013). Updated information is available from Fedoseev et al. (2016). A gold coated substrate is mounted onto the cold tip of a He close-cycle cryostat that is positioned in the center of an UHV main chamber with a typical base pressure of ~$10^{-10}$ mbar. The temperature of the substrate is regulated in the range between 13 and 330 K by means of resistive heating and monitored using two thermocouples. A LakeShore 340 temperature controller allows for a 0.5 K relative temperature precision. Two independently pre-pumped full-metal dosing lines are used for carbon monoxide and glycolaldehyde vapor deposition. The impacting H-atoms are obtained through thermal cracking, using a Hydrogen Atom Beam Source (HABS, Tschersich 2000) or through dissociation of $H_2$ molecules in a capacitively coupled microwave discharge using a Microwave Atom Source (MWAS, Anton et al. 2000). The two sources are mounted in separate UHV chambers that are connected to the main chamber *via* UHV shutters. In both systems the produced H atoms and undissociated $H_2$ molecules remaining in the H-atom beam are thermalized through collisions inside a bent quartz pipe prior to their impact on the ice substrate in the main chamber. This setting allows a fully independent operation of the individual deposition lines. The use of two different H-atom sources also allows for cross-reference of the obtained results. Both HABS and MWAS are commercial systems. Their fluxes are calibrated '*in situ*' and the details of this procedure are available from Ioppolo et al. (2013). Typical H-atom fluxes used in the present study amount to $8\times10^{12}$ atoms cm$^{-2}$ s$^{-1}$.

Along the proposed reaction scheme a steady abundance decrease is expected with increasing COM size. Previously obtained experimental results show that hydrogenation of pure CO ice yields glycolaldehyde and ethylene glycol, but their final abundances were at the edge of the detection sensitivity. This will prohibit the detection of glyceraldehyde and glycerol in a similar setting. Therefore, here the glycolaldehyde content is artificially increased with respect to the amounts formed in the pure CO hydrogenation experiments. Our assumption is that if the formation yield of glycerol and glyceraldehyde from glycolaldehyde is comparable to that of glycolaldehyde and ethylene glycol starting from CO hydrogenation - $HOCH_2CHO/H_2CO$ and $HOCH_2CH_2OH/CH_3OH$ ratios of ~0.01-0.09 were found in Fedoseev et al. 2015 - then the laboratory detection of three-carbon bearing molecules should be possible. Although this approach greatly enhances the relative yields of newly formed species with respect to the initial amount of CO, the yields of three-carbon bearing COMs with respect to glycolaldehyde should not be significantly affected, and the successful detection of three-carbon bearing COMs should validate the overall mechanism proposed here.

The experiments are performed using a 'co-deposition' technique. In this case, CO and $HOCH_2CHO$ molecules as well as the H-atoms are exposed to a clean and precooled substrate resulting in the growth of a uniform ice. Typically ~$1.4 \times 10^{16}$ molecules $cm^{-2}$ are deposited over the duration of a 360 min co-deposition with a 1:1 ratio. The ice composition and the column densities are monitored *in situ* by means of Reflection Absorption Infrared (RAIR) spectroscopy. The band strength (cm $mol^{-1}$) of CO, $H_2CO$, $CH_3OH$ and $HOCH_2CHO$ signals are derived using laser (632.8 nm) interference methods as previously described in Baratta & Palumbo (1998) and Paardekooper et al. (2016b). The values derived for the pure ices are then applied to the ice mixtures. In the experiments $^{13}CO$ (Sigma-Aldrich, 99 % $^{13}C$, <5 % $^{18}O$ isotope), $^{13}C^{18}O$ (Sigma-Aldrich, 99 % $^{13}C$ and 99 % $^{18}O$ isotope), regular CO (Linde 2.0) and $H_2$ (Praxair 5.0) are used as gas supplies. The glycolaldehyde vapors are obtained by thermal decomposition of solid glycolaldehyde dimers (Sigma-Aldrich) under vacuum at ~80 K.

Upon completion of the co-deposition experiment, a temperature programmed desorption (TPD) measurement is performed, typically with a 5 K/min rate. Sublimating species are recorded in the gas phase, as a function of the ice temperature, using a quadrupole mass-spectrometer (QMS), applying 70 eV, *i.e.* 'hard', electron ionization. Dissociative electron impact ionization occurring in the ion source of the QMS causes significant fragmentation of the parent species. The produced QMS spectra are compared with those available in the NIST database[1] for similar electron energies. This provides an additional diagnostic tool to the regular assignment based on the sublimation temperature of individual species. Dissociative ionization also comes with the advantage that it allows to study isotope labeled experiments, using $^{13}CO$ and $^{13}C^{18}O$, to further constrain assignments by searching for the corresponding shifts in the mass spectra. In addition to the QMS measurements, sublimating ice species can also be monitored by acquiring RAIR difference spectra prior and after desorption of each of the ice constituents.

All experiments have been performed using both HABS and MWAS and produce similar results. In the following only the HABS results are presented as these have a lower content of background $H_2O$ contamination. The chosen experiments are also repeated for faster TPD rates of 20 and 25 K/min to obtain higher signal-to-noise ratios for the COM signals. In the latter case, a regular 5 K/min rate is used

---

[1] NIST Mass Spec Data Center, S. E. Stein, director, "Mass Spectra" in NIST Chemistry WebBook, NIST Standard Reference Database Number 69, Eds. P.J. Linstrom and W.G. Mallard, National Institute of Standards and Technology, Gaithersburg MD, 20899

in the range from 15 to 50 K to remove CO ice and, subsequently, 20 or 25 K/min rates are applied above 50 K.

## 4. RESULTS AND DISCUSSION

In **Figure 2** a typical example of a RAIR spectrum is presented, obtained after the co-deposition of HOCH$_2$CHO and CO molecules with H atoms at 15 K. This spectrum is normalized to 5x10$^{15}$ deposited molecules and compared with five reference spectra recorded for pure HOCH$_2$CHO, HOCH$_2$CH$_2$OH, CO, H$_2$CO and CH$_3$OH ices. The latter five spectra are each normalized to 1x10$^{15}$ molecules. Along with the RAIR absorption features of the deposited HOCH$_2$CHO and CO molecules, the absorption bands of the three main hydrogenation products - HOCH$_2$CH$_2$OH, H$_2$CO and CH$_3$OH - can be identified. As discussed in **sections 2** and **3** the detection of the three-carbon bearing sugar and sugar alcohol homologues, i.e., glycerol and glyceraldehyde, cannot be confirmed based solely on the RAIR data due to spectral overlap with IR features of the two-carbon bearing homologues and the relatively low formation yields. Below we address glycerol and glyceraldehyde formation based on the obtained QMS TPD data.

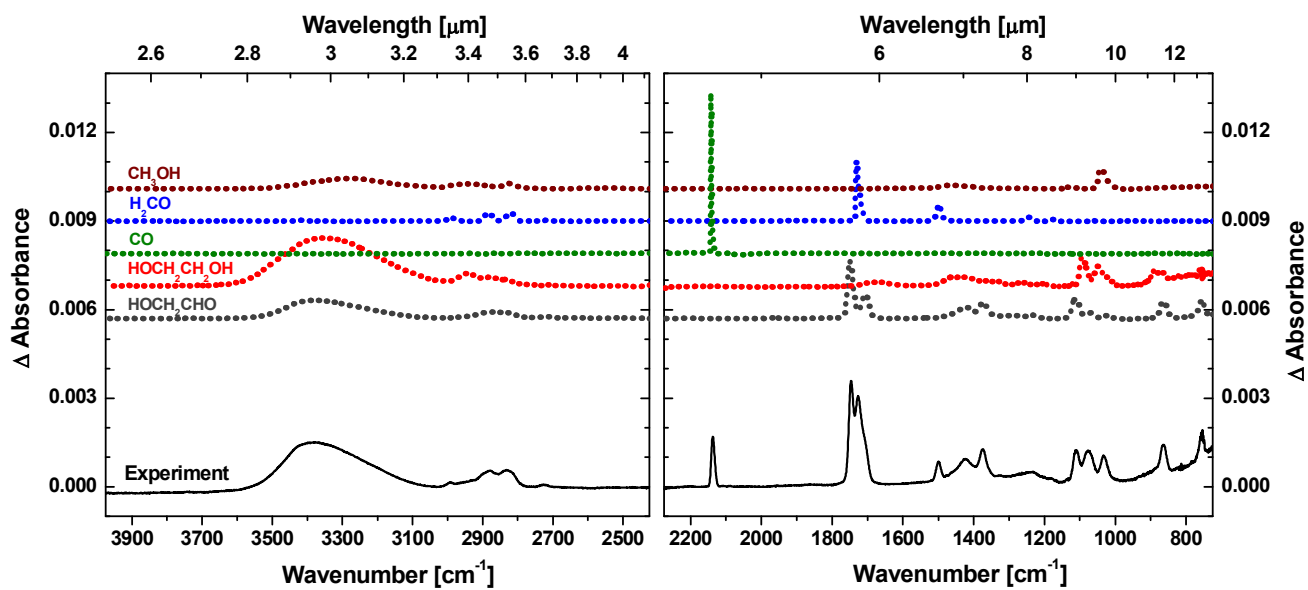

**Figure 2.** Solid line: RAIR spectra obtained after the co-deposition of a CO:HOCH$_2$CHO = 1:1 ice mixture with H-atoms at 15 K. The total CO and HOCH$_2$CHO fluences amount to 7x10$^{15}$ molecules cm$^{-2}$, while the total H-atom fluence is 2x10$^{17}$ atoms cm$^{-2}$. Δ Absorbance is normalized to 5x10$^{15}$ deposited molecules. Dashed lines: the reference spectra obtained after the deposition of pure HOCH$_2$CHO, HOCH$_2$CH$_2$OH, CO, H$_2$CO and CH$_3$OH ices at 15 K. Each spectrum is normalized to 1x10$^{15}$ deposited molecules.

**Glycerol.** For the UHV conditions in SURFRESIDE$^2$, a desorption temperature of glycerol of around 240 K is expected (Kaiser, Maity & Jones 2015; Paardekooper et al. 2016a). All recorded TPD QMS spectra, obtained after simultaneous deposition of HOCH$_2$CHO with CO molecules and H atoms, reveal the presence of a desorption peak at this temperature. The following evidence of glycerol assignment to this desorption feature can be provided.

In the left panel of **Figure 3** an example of a typical TPD QMS spectrum is shown for a temperature range from 60 to 260 K. This spectrum is recorded after co-deposition at 15 K of HOCH$_2$CHO and CO molecules with H atoms for 360 minutes, where molecular deposition rates and H-atom flux amount to $3 \times 10^{11}$ molecules cm$^{-2}$ s$^{-1}$ and $8 \times 10^{12}$ atoms cm$^{-2}$ s$^{-1}$, respectively. Five selected m/z values are plotted in the figure. These are the three characteristic QMS peaks of glycerol, *i.e.*, 31 - CH$_2$OH$^+$, 61 – HOCH$_2$CHOH$^+$ and 43 - C$_2$H$_2$OH$^+$. The two other m/z values at 62 and 64 are shown for comparison. The glycerol QMS spectrum does not reveal any signal at m/z = 92 – the 'parent' molecular mass (NIST database; Kaiser, Maity & Jones 2015), reflecting near complete dissociative ionization of glycerol upon electron impact in the ion source of the QMS. Among the produced characteristic ions, CH$_2$OH$^+$ (m/z = 31) and HOCH$_2$CHOH$^+$ (m/z = 61) are obtained through the cleavage of C-C bonds in glycerol molecules. The m/z = 43 signal (C$_2$H$_2$OH$^+$) is the result of the subsequent loss of a H$_2$O unit from HOCH$_2$CHOH$^+$ ion. The m/z signals at 61 and 43 clearly peak around 240 K, while the m/z = 31 signal is to a large extent obscured by the high baseline value that is caused by species desorbing at lower temperatures.

Besides the 240 K peak, four other distinct desorption features are observed in this diagram. Two peaks can be assigned to well-known CO hydrogenation products, *i.e.,* H$_2$CO at ~100 K and CH$_3$OH at ~140 K (see Hiraoka et al. 1994; Watanabe & Kouchi 2002; Fuchs et al. 2009). Note that the m/z value at 31 corresponds to the parent mass of naturally occurring H$_2^{13}$CO. The intensity of the regular H$_2$CO isotope (m/z = 30, not shown here) is about two orders of magnitude higher, in accordance with the $^{12}$C/$^{13}$C natural abundance ratio. Two other desorption peaks, at ~160 K and ~200 K, can be respectively assigned to unprocessed glycolaldehyde, left over after co-deposition, and to HOCH$_2$CH$_2$OH, a known reaction product upon HOCH$_2$CHO hydrogenation (Fedoseev et al. 2015; Burke et al. 2015; Maity et al. 2015; Chuang et al. 2016). The detection of the formed H$_2$CO, CH$_3$OH and HOCH$_2$CH$_2$OH in addition to the low-volatile product desorbing at 240 K is fully consistent with the reaction scheme presented in **Figure 1**.

To further constrain the assignment of the glycerol desorption peak at 240 K, isotope labelled experiments are performed. The co-deposition of HOCH$_2$CHO with CO molecules and H atoms at 15 K is repeated for $^{13}$CO and $^{13}$C$^{18}$O. In the case that CO-addition occurs at the edge of a HOCH$_2$CHO molecule, see **Figure 1**, isotopically enriched glycerol molecules are formed; $HOCH_2CH(OH)^{13}CH_2OH$ and $HOCH_2CH(OH)^{13}CH_2^{18}OH$ for $^{13}CO$ and $^{13}C^{18}O$, respectively. To our knowledge, the electron impact ionization induced QMS spectra of these two isotopologues have not been reported before, but it is possible to deduce these from the accessible QMS spectra of regular $^{12}C$-containing species (NIST database; Kaiser, Maity & Jones 2015) in the following way.

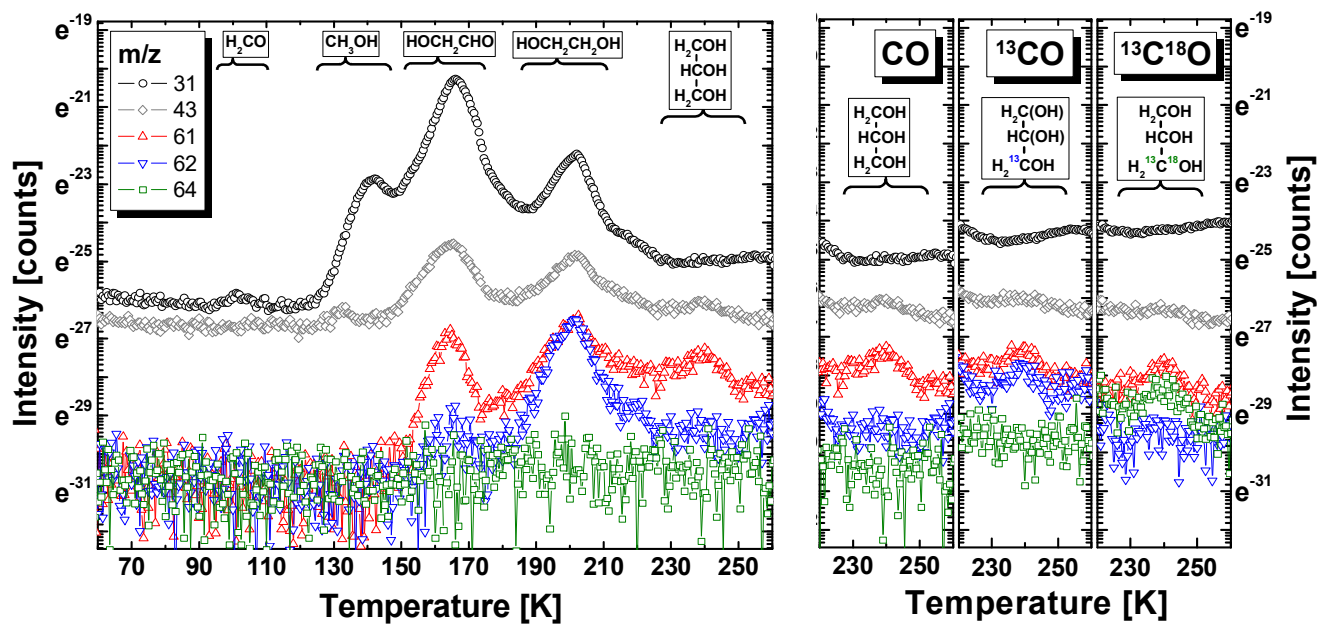

**Figure 3.** Left panel: QMS TPD spectra in the range from 60 to 260 K for five selected m/z values obtained after the co-deposition of a $CO:HOCH_2CHO = 1:1$ ice mixture with H-atoms at 15 K. Right panel: comparison between the fragments of three QMS TPD spectra in the range from 220 to 260 K. The spectra are obtained after co-deposition at 15 K of $HC(O)CH_2OH$ molecules with H-atoms and naturally abundant CO (left spectrum), $^{13}CO$ (central spectrum) or $^{13}C^{18}O$ (right spectrum) labeled isotopes. All presented experiments are performed for identical experimental conditions. The total CO ($^{13}CO$, $^{13}C^{18}O$) and $HOCH_2CHO$ fluences amount to $7\times10^{15}$ molecules cm$^{-2}$, while the total H-atom fluence is $2\times10^{17}$ atoms cm$^{-2}$. The ordinate axis is presented in the natural logarithm scale for clarity.

Glycerol is a symmetric molecule. Assuming that each of the symmetric chemical bonds has an equal probability to break upon 70 eV electron impact, the resulting distribution of m/z intensities can be interpreted. Take as an example the cleavage of C-C bonds of glycerol. As previously discussed, two of the most intense m/z values in the QMS spectrum of glycerol are 31 ($CH_2OH^+$) and 61 ($HOCH_2CHOH^+$). In the case that only a single edge carbon or oxygen atom of the glycerol molecule is labeled by the isotope, half of the produced $CH_2OH^+$ ions should contain this isotope and half should not. On the contrary, in the case that a central carbon or oxygen atom in the glycerol molecule is $^{13}C$ or $^{18}O$ labeled, no drop in m/z = 31 signal is expected. A similar conclusion can be deduced for the $HOCH_2CHOH^+$ ion where the same 1-to-1 distribution between isotopically labeled and unlabeled fragments should be obtained when the edge carbon or oxygen atom is isotopically labeled. In the case that the central carbon or oxygen is labeled, all resulting $HOCH_2CHOH^+$ ions will contain isotopically labeled atoms. Therefore, under the fair assumption that the dissociative ionization cross-sections are nearly identical for isotopically labeled and non-labeled glycerol, a 50% intensity drop at m/z = 61 is expected. The corresponding intensity increase at m/z = 62 or 64 should be observed for $HOCH_2CH(OH)^{13}CH_2OH$ or $HOCH_2CH(OH)^{13}CH_2^{18}OH$ labeled glycerol, respectively. These experiments have been performed and the resulting TPD spectra obtained after co-deposition of $HOCH_2CHO$ with H atoms and CO, $^{13}CO$ or $^{13}C^{18}O$ molecules are shown in the three right panels of **Figure 3** in the range from 220 to 260 K. The measurements exhibit the predicted intensity shifts for m/z = 61, 62 and 64. None of the m/z = 62 and m/z = 64 signals exhibit a peak at 240 K for the regular glycerol isotope, whereas in the case of $^{13}C$ labelled carbon monoxide, two peaks can be observed for m/z = 61 and m/z = 62 corresponding to the $HOCH_2CHOH^+$ and $HO^{13}CH_2CHOH^+$, respectively. In the case of $^{13}C^{18}O$, m/z = 61 ($HOCH_2CHOH^+$) and m/z = 64 ($H^{18}O^{13}CH_2CHOH^+$) peaks are observed while no rise of m/z = 62 signals is detected.

By extrapolating the procedure described above to other ions observed in the QMS spectra of glycerol, the corresponding $HOCH_2CH(OH)^{13}CH_2OH$ and $HOCH_2CH(OH)^{13}CH_2^{18}OH$ spectra can be evaluated, from the natural 98.9 % $^{12}C$ isotope QMS spectrum of glycerol available in the NIST database. The obtained QMS reference spectra are presented in the two upper panels of **Figure 4**. Two m/z ranges are chosen. On the left panel the previously discussed range between m/z = 60 and m/z = 64 is presented, while in the right panel the range between m/z = 43 and m/z = 47 is shown. In the two lower panels, the experimentally obtained intensities are presented for comparison. A good agreement between the evaluated QMS spectra of the different glycerol isotopes and the m/z intensities obtained from our experiments is found. A conclusive fit is obtained for the m/z = 60 to 64 range, while for the m/z = 43 to 47 range some discrepancies are observed. These are expected, however. Most of the ions in this range are formed by secondary fragmentation, *e.g.,* $H_2CCOH^+$ (m/z = 43) is obtained by the loss of $H_2O$ unit from $HOCH_2CHOH^+$ (m/z = 61). The latter ion lacks symmetry. Therefore, a deviation from a statistical 1-to-1 ratio is expected. Furthermore, some of the m/z signals can be caused by more than one ion, *e.g.* $HOCH_2CH^+$, $HOCHCH_2^+$, $CO_2^+$ *etc.*, all have m/z = 44.

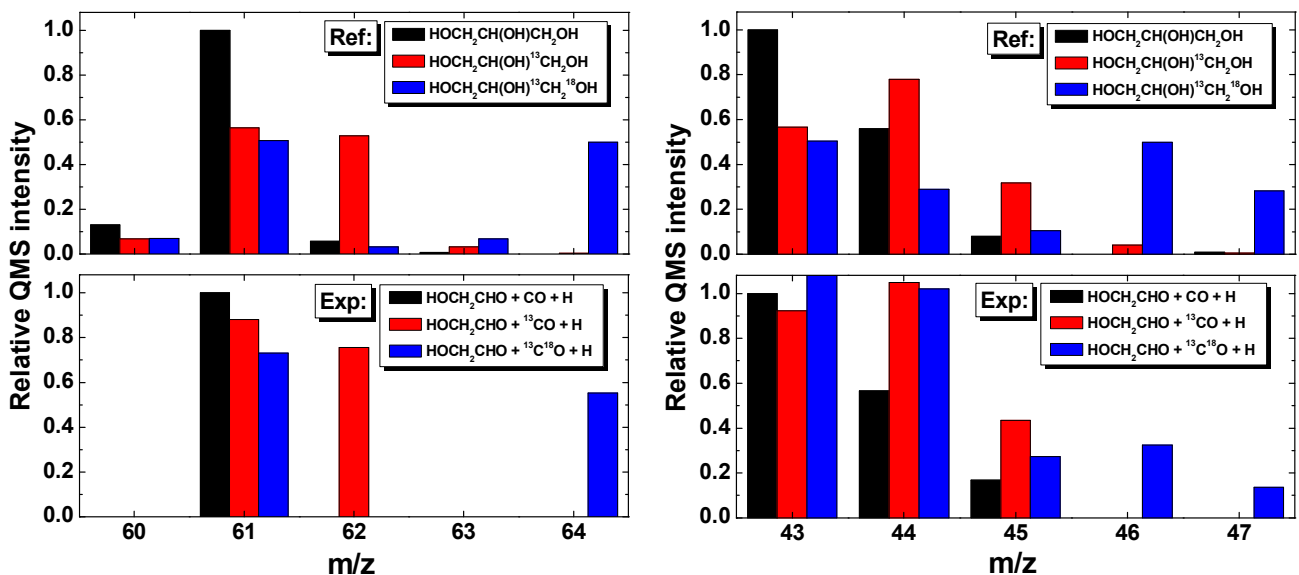

**Figure 4.** Left-top panel: the 70 eV QMS spectra of three distinct glycerol isotopes $HOCH_2CH(OH)CH_2OH$ (NIST database), and $HOCH_2CH(OH)^{13}CH_2OH$ with $HOCH_2CH(OH)^{13}CH_2^{18}OH$ (evaluated from the NIST database data in this study). The range from m/z = 60 to m/z =64 is shown. Left-bottom panel: comparison with the QMS intensities obtained at 240 K (lower panel) after the 15 K co-deposition of $HOCH_2CHO$ molecules with H-atoms and naturally abundant CO (black bars), $^{13}CO$ (red bars) or $^{13}C^{18}O$ (blue bars) labeled isotopes. The total CO ($^{13}CO$, $^{13}C^{18}O$) and $HOCH_2CHO$ fluences are equal to $7\times10^{15}$ molecules $cm^{-2}$, while the total H-atom fluence amounts to $2\times10^{17}$ atoms $cm^{-2}$. Right panels: The same spectra in the range from m/z = 43 to m/z =47, normalized to the m/z = 43 signal in a regular CO experiment.

Based on the fit between the predicted and experimentally obtained glycerol QMS spectra presented in **Figure 4**, the formation of glycerol in our experiments is concluded. Moreover, the obtained fit confirms that CO addition occurs at the edge of the $HOCH_2CHO$ molecules, constraining the mechanism shown in **Figure 1**. Assuming similar electron impact ionization cross-sections and pumping rates of ethylene glycol (the two-carbon bearing product of glycolaldehyde hydrogenation), and of glycerol (the three-carbon bearing product) $HOCH_2CH(OH)CH_2OH/HOCH_2CH_2OH$ ratio of about 0.01 is derived from the TPD spectra presented in **Figure 2**. This indicates that for our experimental conditions, only a limited fraction of the $HOCH_2CHO$ molecules contributes to a further extension of the carbon skeleton, as discussed in Sections 3 of this work.

**Glyceraldehyde.** QMS spectra of isotopically labelled $HOCH_2CH(OH)^{13}CHO$ and $HOCH_2CH(OH)^{13}CH^{18}O$ cannot be extrapolated in a similar way as in the case of glycerol, due to the lack of molecular symmetry of glyceraldehyde. Nevertheless, several experimental observations have been made. Two desorption temperatures have been reported for pure glyceraldehyde under UHV conditions (McManus 2014). These are 215 and 280 K, and these two desorption peaks were assigned to glyceraldehyde monomers and dimers in the adsorbed multilayers, respectively. **Figure 5** shows the fragments of three TPD spectra in the 202 to 235 K temperature range where thermal desorption of glyceraldehyde monomers is expected. These mass spectra are recorded after the co-deposition of $HOCH_2CHO$ with H atoms and CO, $^{13}CO$ or $^{13}C^{18}O$ molecules, respectively, discussed in the glycerol section. The results for the same m/z values as used before, *i.e.,* 31, 43, 61, 62 and 64, are shown, where m/z = 31, 43 and 61 correspond to the most intensive peaks in the glyceraldehyde mass spectrum (NIST database; Brittain et al. 1971). For clarity, the ordinate axis is depicted as a natural logarithm scale, where exponential decay appears as a straight line. A clear deviation from a straight decay starting from 210 K for m/z = 31 and m/z = 43 (and possibly m/z = 61) plots is observed indicating the appearance of a new species. In this case both the desorption temperature and the affected m/z values are in line with the desorption of glyceraldehyde monomers. Further evidence constraining the desorption of new species between 215 and 220 K is obtained in the $^{13}CO$ labeled experiments, in which a clear desorption peak for m/z = 62 is observed, while in the $^{13}C^{18}O$ labeled experiment a peak at m/z = 64 may be present. These observations point at the formation of glyceraldehyde in the performed experiments However, an unambiguous assignment, as in the case of glycerol, is not possible at this stage and requires the use of more sensitive or selective techniques (Maity, Kaiser & Jones 2015, Paardekooper et al. 2016a; Oba et al. 2016). However, given the positive identification of glycerol, following the reaction scheme shown in **Figure 1**, formation of glyceraldehyde is plausible with the observations presented here.

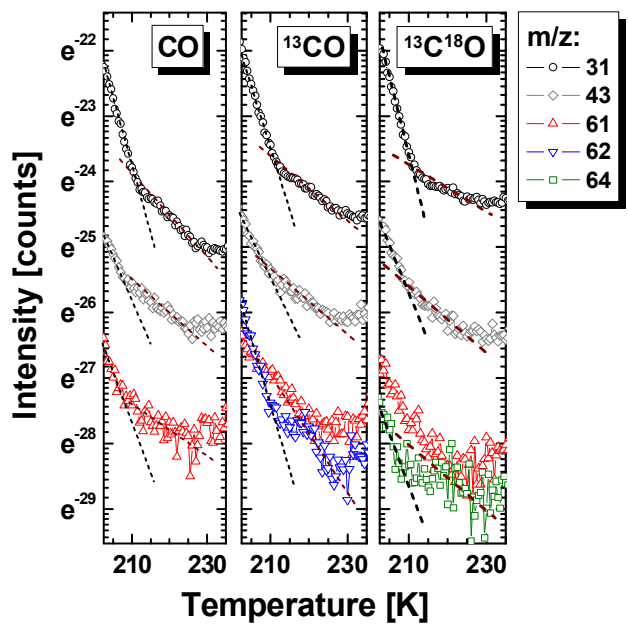

**Figure 5**. Comparison between the fragments of three QMS TPD spectra in the range from 202 to 235 K. The spectra are obtained after 15 K co-deposition of $HOCH_2CHO$ molecules with H-atoms and regular CO (left spectrum), $^{13}CO$ (central spectrum) or $^{13}C^{18}O$ (right spectrum) labeled isotopes. The total CO ($^{13}CO$, $^{13}C^{18}O$) and $HOCH_2CHO$ fluences amount to $7x10^{15}$ molecules $cm^{-2}$. The total H-atom fluence amounts to $2x10^{17}$ atoms $cm^{-2}$. The ordinate axis is presented as a natural logarithm and an exponential decay should appear as a straight line. Dashed lines are added to highlight the presence of a distinct source of ions appearing at ~210 K, consistent with glyceraldehyde desorption.

## 5. CONCLUSIONS AND ASTRONOMICAL RELEVANCE

Our laboratory results successfully demonstrate that in accordance with the reaction scheme presented in **Figure 1** three-carbon bearing COMs can be formed starting from simple accretion of H atoms and CO molecules on the surface of interstellar grains. Conclusive evidence for glycerol formation has been obtained in a set of selected isotope labeled experiments. The formation of glyceraldehyde is shown to be plausible. The reaction scheme presented in **Figure 1** is the extension of the well studied CO + H hydrogenation chain resulting in the formation of $H_2CO$ and $CH_3OH$ ice. This solid state reaction chain is included in many astrochemical models, to explain solid and gas-phase $CH_3OH$ observations in dark clouds and young stellar objects. Thus, the reaction routes investigated in this study can be used to extend already existing astrochemical simulations. In this respect, the lattice-gas kinetic Monte Carlo simulations are the most promising as they account for the specific lattice position for each of the intermediate radicals required to form COMs with two, three and more carbon atoms (Cuppen et al. 2013, Vasyunin & Herbst 2013a, Chang & Herbst et al. 2016, Garrod et al. 2017). However, the direct implications of the formation routes presented in **Figure 1** requires better knowledge of branching ratios and the activation barriers of various H-atom addition and abstraction reactions. Moreover, the correct implementation of quantum tunneling rate constants is of importance as these also affect branching ratios for various H-atom addition and abstraction reactions (Lamberts et al. 2016). These issues can be partially overcome by extrapolating branching ratios and activation barriers from similar reactions investigated in previous studies (Fedoseev et al. 2015). In the latter case, the authors employed a model previously used to fit laboratory data of $H_2CO$ and $CH_3OH$ formation (Fuchs et al. 2009) and extrapolate the results to astrochemical time-scales. H-atom abstraction reactions were omitted, and the effective reaction activation barrier for H-atom addition to the aldehyde group of $HOCH_2CHO$ was assumed to be equal to the effective reaction activation barrier of H-atom addition to $H_2CO$ as obtained in Fuchs et al. 2009. In a similar way, a first approximation could be made for H-atom addition to the aldehyde group of $HOCH_2CH(OH)CHO$. The R-ĊHOH/R-$CH_2O$· branching ratios obtained in these H-atom addition reactions can be set to 0.5 or 1 to provide estimates from above for the amount of formed COMs.

With increasing carbon chain length, the expected COM yields reduce strongly for each of the newly added –C(O)- segments. This complicates the assignment of larger species in the experiments starting from pure CO hydrogenation. The ice samples enriched with two-carbon bearing precursors are used to confirm formation of the three-carbon bearing COMs, and further exploit the reaction scheme shown in **Figure 1**. In addition to the experimentally obtained $HOCH_2CH_2OH/CH_3OH$ ratio of ~0.01-0.09 previously reported by Fedoseev et al. (2015), a $HOCH_2CH(OH)CH_2OH/HOCH_2CH_2OH$ ratio of ~0.01 is reported in present study. It should be stressed, however, that this value is obtained under strong assumptions that include similar ionization cross-sections and pumping rates of both molecules. Astrochemical simulations are needed to predict astronomically relevant abundance ratios.

Glycolaldehyde and ethylene glycol have been observed in solar-mass protostars. The present study shows that reactions of H and CO with glycolaldehyde can lead to large oxygen-rich COMs in cold dark clouds, *i.e.*, without the need of energetic processing of the ice mantle. Instead, 'non-energetic' H-atom addition and abstraction reactions provide the reactive intermediates to generate glycerol and likely glyceraldehyde. So far, similar results for oxygen-rich three-carbon bearing COMs were only presented after VUV-photon or cosmic ray irradiation of $CH_3OH$ containing ices, where various reactive intermediates, *i.e.,* ·$CH_2OH$, $CH_3O$·, ·$CH_3$, HĊO, *etc*. are obtained by $CH_3OH$ dissociation with impinging particles (Kaiser, Maity & Jones 2015; Maity, Kaiser & Jones 2015; Paardekooper et al. 2016a). The latter process is effective but lacks the selectivity of an H-atom induced reaction scheme as shown in **Figure 1**. Furthermore, H-atom induced reactions do not cause newly formed species to fragment, in contrast to impinging VUV-photons or cosmic rays. It should be stressed, however, that the formation path for simple sugars and sugar alcohols as presented here does not contradict with the 'energetic route' suggested in these studies. It rather complements these, showing that COMs of astrobiological importance can be formed as early as the cold dark cloud stage, during the CO freeze out stage and well before formation of the protostar and heating of the dust occurs. COMs produced on icy dust grains in this early stage of star formation can, in turn, be locked on the grain surface, and participate in active energetic processing during later stages, or even be released into the gas phase by shocks, heat or photodesorption (Charnley et al. 2008; Herbst & van Dishoeck 2009; Caselli & Ceccarelli 2012; Walsh et al. 2014a,b, Jimenez-Serra et al. 2016). Their efficient release into the gas-phase by heat and photodesorption is unlikely, however, due to the low-volatility of such complex species, and a tendency of even the simplest COMs (*e.g.* $CH_3OH$) to photodissociate and desorb as fragments, rather than as an intact molecule (Bertin et al. 2016). As a consequence, these heavy weight COMs have a higher chance to remain on grain surfaces during the later stages of ice evolution.

The astronomical importance of our experimental findings is emphasized by the astrobiological role of the obtained species. Glycerol is a necessary component of phospholipids (consisting of fatty acids, glycerol and inorganic phosphate) comprising the membranes of living cells. Glyceraldehyde is the simplest sugar that plays a key role in the energy transfer inside the cells of living organisms. Furthermore, the formation of glycerol and glyceraldehyde along the steps discussed here means that even larger sugar and sugar alcohols are expected to be formed by hydrogenation of accreting CO molecules, as shown in **Figure 1**. Thus, the presence of simple sugars and sugar alcohols on young planets is possible under the assumption that at least a fraction of the original icy-dust material survives upon transfer to the early planet surface or, alternatively, is delivered by comets or other celestial bodies during the late bombardment stage of the early Earth. From this point of view, it should be noticed that the reaction scheme suggested in **Figure 1** have to result in the formation of racemic mixtures of simple sugars and sugar alcohols, and the selection of D- over L-stereoisomers should occur at one of the later evolutionary stages.

**Acknowledgement**

This research was funded through a VICI grant of NWO (the Netherlands Organization for Scientific Research) and A-ERC grant 291141 CHEMPLAN. The financial support by NOVA (the Netherlands Research School for Astronomy) and the Royal Netherlands Academy of Arts and Sciences (KNAW) through a professor prize is acknowledged. GF acknowledges the financial support from the European Union's Horizon 2020 research and innovation programme under the Marie Sklodowska-Curie grant agreement n. 664931. SI acknowledges the Royal Society for financial support and the Holland Research

School for Molecular Spectroscopy (HRSMC) for a travel grant. The described work has benefitted a lot from collaborations within the framework of the FP7 ITN LASSIE consortium (GA238258).